\documentclass[twocolumn,epjc3]{svjour3}          

\RequirePackage[T1]{fontenc}
\usepackage{amssymb}
\usepackage{amsmath}
\usepackage{bm}
\smartqed  
\newcommand{\vf}{v_\text{F}}

\newcommand{\ee}{\text{e}}
\newcommand{\ii}{\text{i}}

\RequirePackage{graphicx}
\RequirePackage{flushend}
\usepackage[usenames,dvipsnames]{xcolor}%
\definecolor{goodgreen}{rgb}{0.1,0.5,0}
\definecolor{goodred}{rgb}{0.7,0,0}
\definecolor{tinekegreen}{RGB}{22,140,22}
\definecolor{ferraro}{rgb}{0.2, 0.5, 0.}

\usepackage[colorlinks,urlcolor=goodgreen,citecolor=blue,linkcolor=goodred]{hyperref}

\usepackage[normalem]{ulem}

\journalname{Eur. Phys. Special Topics}

\begin{document}

\title{Wave-particle duality of electrons with spin-momentum locking}

\subtitle{Double-slit interference and single-slit diffraction effects of electrons on the surface of three-dimensional topological insulators}

\author{Dario Bercioux\thanksref{e1,addr1,addr2}
        \and
        Tineke L. van den Berg\thanksref{e2,addr1} \and Dario Ferraro\thanksref{addr3,addr4,addr5} \and J\'er\^{o}me Rech\thanksref{addr3} \and Thibaut Jonckheere\thanksref{addr3} \and Thierry Martin\thanksref{addr3}  
}

\thankstext{e1}{e-mail: dario.bercioux@dipc.org}
\thankstext{e2}{e-mail: tineke.vandenberg@dipc.org}

\institute{Donostia International Physics Center (DIPC), Manuel de Lardizbal 4, E-20018 San Sebasti\'an, Spain\label{addr1}
          \and
          IKERBASQUE, Basque Foundation of Science, 48011 Bilbao, Basque Country, Spain\label{addr2}
          \and 
          Aix Marseille Univ, Université de Toulon, CNRS, CPT, Marseille, France\label{addr3}
          \and
          Dipartimento di Fisica, Universit\`a di Genova, Via Dodecaneso 33, 16146, Genova, Italy\label{addr4}
          \and
          SPIN-CNR, Via Dodecaneso 33, 16146 Genova, Italy\label{addr5}
}

\date{\today}

\maketitle

\begin{abstract}
We investigate the effects of spin-momentum locking on the interference and diffraction patterns due to a double- or single-slit in an electronic \emph{Gedankenexperiment}.  We show that the inclusion of the spin-degree-of-freedom, when coupled to the motion direction of the carrier --- a typical situation that occurs in systems with spin-orbit interaction --- leads to a modification of the interference and diffraction patterns that depend on the geometrical parameters of the system.
\end{abstract}

\section{Introduction}

The wave-particle duality is one of the fundamental paradigms introduced by quantum mechanics, which tells us that every particle or quantum entity may be described as either a particle or a wave~\cite{eisberg_resnick_2009}. In one of his ``Lectures on physics'' books, Feynman \emph{et al.}~\cite{richardfeynman_2011} proposed to verify the wave nature of electrons by performing a thought experiment analogous to the one conducted by Thomas Young, performed in the first decade of the 1800s to show the wave nature of light. The first experiment implementing the Young experiment with electrons was realized by  J\"onsson~\cite{Joensson_1961,Joensson_1974} contemporaneously with the preparation of Feynman's lecture notes; these results were confirmed a few years later by a team of researchers at the University of Bologna~\cite{Merli_1976}. In this experiment, the two slits of the setup by J\"onsson were substituted by a biprism. Further refinement came after more than a decade with an experiment performed at the Hitachi lab by Tonomura \emph{et al.}~\cite{Tonomura_1989}. The readers of the magazine ``Physics World'' of the Institute of Physics selected these experiments 
to be the most beautiful ones in physics of the past century~\cite{Steeds_2003}. 

Although this type of research has now been in large part delegated to the educational framework~\cite{Malgieri_2017,Sayer_2017}, several groups in recent years tried to push the limits of the understanding of the validity of the wave-particle duality towards large quantum objects and molecules. One of the most complex attempts was realized by considering interference and diffraction of large C$_{60}$ molecu\-les~\cite{Arndt_1999,Nairz_2000,Nairz_2003}. This experiment was a breakthrough in the understanding of the limits of quantum theory because the C$_{60}$ molecule is close to being a classical object when considering its many excited internal degrees-of-freedom and also the large possibility of coupling to the environment that can lead to decoherence effects~\cite{Arndt_1999,Cronin_2009,Hornberger_2012}. 
More recent experiments verified interference effects in more complex molecular aggrega\-tes~\cite{Eibenberger_2013}. The problem of addressing the ``wave-particle duality'' rationally in the framework of complex quantum systems has been recently investigated by Carnio \emph{et al.}~\cite{Carnio_2019}.

In the standard setup for the double-slit experiment, the spin degree-of-freedom of the electrons does not play any role. Some recent experiments focusing on electron beams carrying orbital angular momentum~\cite{Hasegawa_2013} showed the appearance of dislocations in the interference pattern. Here we note a strong analogy with photons carrying an orbital angular momentum; in fact these can be generated by  a diffraction grating containing a dislocation~\cite{Bazhenov_1992}.
Recently, the role of Rashba spin-orbit interaction  (RSOI)~\cite{Bychkov_1984} was investigated in the interference pattern in two-di\-mensional electron gases (2DEGs)~\cite{Shimizu_2020} in a time-de\-pendent fashion. The authors conclude that the RSOI does not affect the interference pattern.However, from other investigations it is known spin-orbit interaction is of importance in various quantum optics phenomena, such as in spin-dependent double refraction and pumping~\cite{Bercioux_2010,Bercioux_2012}.

In this work, we focus on an extreme case of the results presented in Ref.~\cite{Shimizu_2020} by considering the surface states of three-dimensional topological insulators (3DTI). These surface states are described by an effective Hamiltonian similar to the one of a 2DEG with RSOI but with an infinite effective electron mass so that the parabolic kinetic term is zero~\cite{Bardarson_2013}. In 2DEG with RSOI, the surface electron states are characterized by the so-called \emph{spin-momentum locking} (SML), \emph{i.e.}, the spin orientation is strongly connected to the motion direction~\cite{Bercioux_2015}. We study two fundamental mechanisms of interference: one arising from a wave passing through a double-slit opening and one by a single slit opening. In the former case, we assume that the width of each slit is of the same order as the electron Fermi wavelength; in the latter case, we relax this restriction. Conventionally, interference refers to the case of interaction of a few waves, whereas diffraction considers the case of a large number of interacting waves. Nevertheless, they represent the same physical wave mechanisms. Contrary to the results of Ref.~\cite{Shimizu_2020}, we work only with stationary states, because the time-dependent part of the wave-function would lead only to an overall intensity prefactor~\cite{born2019principles}. 
Throughout this work, we compare the case of spinless electrons (SEs), \emph{i.e.}, electrons that are spin-degenerate, with electrons with SML. We find that SML leads to a small but finite correction to the interference and diffraction pattern in comparison to a SE system. These corrections can be of the order of a few per cents and tend to disappear in the so-called far-field limit whereas they are more significant in the opposite near-field limit. 
Our results fall in the framework of the so-called \textit{electron quantum optics}~\cite{Bocquillon_2013} for ballistic chiral conductors.  Extensions to the case of the edge states of two-dimensional topological insulators have already been proposed~\cite{Edge_2013,Inhofer:2013,Hofer:2013,Ferraro_2014,Ferraro_2016}. The present work represents an extension of electron quantum optics to the realm of topological surface states.

The mechanism of modification of the interference patterns we study in this work plays an important role when studying quasi-particle interference via scanning tunneling microscopy/spectroscopy (STM/STS)~\cite{Avraham_2018} of the surface of 3DTI as Bi$_2$Te$_3$, Bi$_2$Se$_3$ or similar topological layered material system when perturbed with normal or magnetic impurities~\cite{Roushan_2009,Sessi_2014}.

The article is organized in the following way:  in Sec.~\ref{spinless_case}, we develop the general formalism for studying interference from a double-slit and then diffraction from a single slit for the case of spinless electrons. In Sec.~\ref{spin_momentum_locking_case}, we perform the same analysis by considering the surface state electrons of 3DTI. We first present the low-energy physics of general 3DTIs, before moving on to the high-energy physics of a more specific case valid for certain 3DTIs. In Sec.~\ref{comparison}, we present a comparison between the different cases and show the presence of a finite correction due to SML. We end the work with conclusions and outlook in Sec.~\ref{conclusions}.


\section{Case of spinless electrons}\label{spinless_case}

We start evaluating the interference and diffraction patterns for SEs in a 2DEG. They are described by the quadratic Hamiltonian: 
%
%
\begin{equation}\label{ham_SL}
    \mathcal{H}_\text{SE}=\frac{\bm{p}^2}{2m^*}+V(\bm{r}),
\end{equation}
%
%
where $m^*$ is the effective electron mass of the 2DEG under investigation. In this Hamiltonian, $V(\bm{r})$ is a two-dimensional potential describing a wall with one or more slits, given by the shape of the red lines in Fig.~\ref{fig:young}. Even though we do not explicitly need this potential in our theory, as we work with point-like sources at the positions of the slits, some important physical considerations must be kept in mind. The wall must have a width $w$ that is larger than the decay length $\xi$ of the wavefunction inside the wall $w > \xi$, so that no tunneling will take place. Further, the edges have to be smooth, at least on the order of the long wavelength approximation~\cite{Bercioux_2010}.

\subsection{Interference from double-slit set-up}

\begin{figure}[!tb]
\begin{center}
\includegraphics[width=\linewidth]{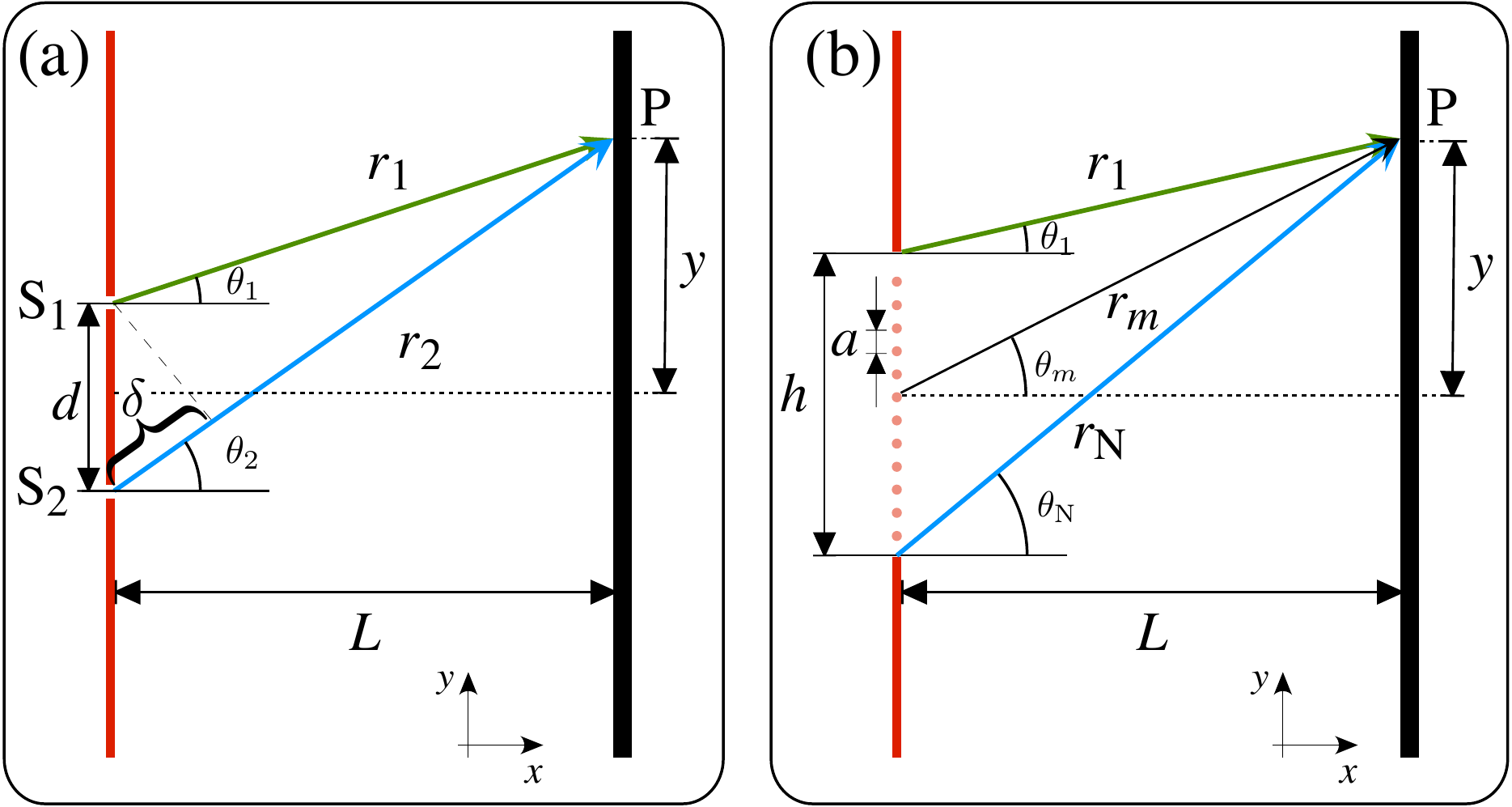}
\caption{\label{fig:young} (a) Sketch of the double-slit set-up for performing the interference experiments for the spinless electrons and for the one with spin-momentum locking. (b) Sketch of the single slit set-up for a diffraction experiment. Here, the opening $h$ is larger than the Fermi wavelength $\lambda_\text{F}$, the points in red in the slit act as secondary wave source and give rise to a diffraction pattern on the screen.}
\end{center}
\end{figure}

We consider first a double-slit setup: these are separated by a distance $d$ and there is a distance $L$ between the slits and an observation plane. Here, we consider the case in which the opening of each single slit $h\sim \lambda_\text{F}$, where $\lambda_\text{F}=2\pi/k_\text{F}$ is the Fermi wavelength associated with the electron Fermi energy $\mathcal{E}_\text{F}=\hbar^2k_\text{F}^2/2m^*$. We will relax this condition when we consider the diffraction pattern in the next section.
The two slits opening $\text{S}_{1/2}$ are the source of two circular plane waves that propagate to the observation point P on the screen at a distance $L$ --- see Fig.~\ref{fig:young}(a). The part of the wave function at the slits that will have an action on the point $P$ on the screen for a given Fermi momentum can be written as:
%
%
\begin{equation}\label{wf_SE}
    \Psi_{1/2}= \ee^{\ii k_\text{F}r_{1/2}(y)}\,,
\end{equation}
%
%
where $r_i$ is the distance between the corresponding slit S$_i$ and the observation point P. From the geometric sketch in Fig.~\ref{fig:young}(a), we have:
%
%
\begin{equation}\label{paths}
r_{1/2}(y)=\overline{\text{S}_{1/2}\text{P}}=\sqrt{L^2+\left(y\mp\frac{d}{2}\right)^2}\,.
\end{equation}
%
%
In the following, we omit writing the explicit $y$ dependence in the $r$ functions.

The intensity at the observation point P is given by the modulus square of the sum of the two wave functions in $y$:
%
%
\begin{subequations}\label{int_SE}
\begin{eqnarray}
    \mathcal{I}_\text{SE}(y)& = & |\Psi_\text{tot}^\text{int}|^2 = |\Psi_1+\Psi_2|^2 \nonumber \\
    & =& 2\left\{1+\cos\left\{k_\text{F}[r_1-r_2]\right\}\right\} \label{int_SE_b} \\
    &=& 4 \cos^2\left\{\frac{k_\text{F}}{2}[r_1-r_2]\right\}\,. \label{int_SE_c}
\end{eqnarray}
\end{subequations}
%
%
The quantity $\Delta r=r_1-r_2$ represents the difference in the geometrical paths between the electron travelling from S$_1$ to P and from S$_2$ to P. 
%
%
%
%
For the difference between the two geometric paths we find:
%
%
\begin{subequations}\label{deltar}
\begin{eqnarray}
    \Delta r=r_1-r_2& =& -\frac{2 y d}{r_1+r_2} \label{deltar_a} \\
    &\simeq & -\frac{y d}{L} \label{deltar_b}
\end{eqnarray}
\end{subequations}
%
%
The approximation in Eq.~\eqref{deltar_b} is valid in the limit $L>d$ and in classical optics it corresponds to the far-field approximation~\cite{born2019principles}. Throughout the manuscript, we will always present results using the definition in Eq.~\eqref{deltar_a}.

\subsection{Diffraction from a single slit}

In this section we relax the condition  $h\sim\lambda_\text{F}$ and assume that the opening of the slit $h$ is larger than the Fermi wavelength $\lambda_\text{F}$. Under this assumption, there are $N=\text{int}[h/\lambda_\text{F}]$ points at a distance $a=\lambda_\text{F}$, --- see Fig.~\ref{fig:young}(b). Each of these points behaves as a secondary emitter of an electron wave, analogously to the Huygens–Fresnel principle in optics~\cite{born2019principles}. 
Under this assumption, the global wave function at the single slit $\Psi_\text{tot}^\text{dif}$ that will have an action on a point $P$ on the screen can be written as:
%
%
\begin{equation}
    \Psi_\text{tot}^\text{dif}=\sum_{m=1}^N \ee^{\ii k_\text{F}r_m}\,,
\end{equation}
%
%
where $r_m$ is the path between the secondary emitter and the observation point P --- it is expressed by a generalization of Eq.~\eqref{paths}. The  intensity $\mathcal{D}_\text{SE}$ at the observation point $y$ is given by
%
%
\begin{eqnarray}\label{diff_SE}
    \mathcal{D}_\text{SE}(y)&=&|\Psi_\text{tot}^\text{dif}|^2 \nonumber \\
    &=&N+\sum_{\substack{m,n=1\\n\neq m}}^N \cos\left[k_\text{F}(r_n-r_m)\right]\,.
\end{eqnarray}
%
%

We note in passing that if the observation plane is placed at very large distance $L\gg h$ and for $y\gg \lambda_\text{F}$, we can assume that the phase difference is $k_\text{F}(r_n-r_m)\to (n-m)\varphi$ , the expression for the diffraction pattern can be simplified to the following well known expression for the optical case~\cite{born2019principles}:
%
%
\begin{equation}\label{far_field_diff_SE}
    \mathcal{D}_\text{SE}= \left[ \frac{\sin\left(N \frac{\varphi}{2}\right)}{\sin\left(\frac{\varphi}{2}\right)}\right]^2
\end{equation}
%
%
where in leading order in $L$, we have 
%
%
\begin{equation}\label{defvarphi}
    \varphi=\frac{2\pi}{\lambda_\text{F}}\frac{ay}{L}.
\end{equation}
%
%

\section{Case of spin-momentum locked electrons}\label{spin_momentum_locking_case}


In this section we consider the surface states of a 3DTI described by the following Hamiltonian
%
%
\begin{equation}\label{ham:3DTI}
\mathcal{H}_\text{SML}= \mathcal{H}_\text{3DTI} +V(\bm{r})\sigma_z\,,
\end{equation}
%
%
where $\mathcal{H}_\text{3DTI}$ is the term describing the surface states of the 3DTI and $V(\bm{r})$ is again an opportune two-dimensional potential describing the single or double-slit setup --- see Fig.~\ref{fig:young}. The potential $V(\mathbf{r})$ is  proportional to the Pauli matrix $\sigma_z$ so as to open a gap in  the linear dispersion of the surface states locally in space. In the following we will first concentrate on the low-energy physics, with a Hamiltonian describing any generic 3DTI. We will then move on to the physics at higher energies and include the hexagonal warping of the Fermi surface, as happens at the surface of 3DTIs with a rhombohedral crystalline structure. 

\subsection{Low-energy case for the 3DTI}\label{section_low_energy}

In this section, we will consider the low-energy approximation Hamiltonian describing the surface states of a generic 3DTI, this reads:
%
%
\begin{equation}\label{eq_low_energy}
    \mathcal{H}_\text{3DTI}=v_\text{F}\left(\bm{\sigma} \times \bm{p}\right)_z.
\end{equation}
%
%
This Hamiltonian is characterized by a linear spectrum $E_\pm(k)=\pm\hbar \vf \kappa$ with $\kappa=\sqrt{k_x^2+k_y^2}$, and eigenstates
%
%
\begin{equation}\label{eq_ei_SML_LE}
    v_\pm=\frac{1}{\sqrt{2}}\begin{pmatrix}1 \\
    \mp \ii\ee^{\ii\theta}
    \end{pmatrix},
\end{equation}
%
%
where $\theta=\arctan(k_y/k_x)$. It is worth noticing that these eigenstates are identical to the ones that can be obtained for a 2DEG with RSOI~\cite{Bercioux_2015}. 

\subsubsection{Interference from double-slit set-up}

In order to study the interference pattern,
we consider two plane waves with a spinorial structure as introduced in Eq.~\eqref{eq_ei_SML_LE}:
%
%
\begin{equation} \label{wavesint1}
\tilde{\Psi}_{1/2}(y)=\frac{\ee^{\ii k_\text{F} r_{1/2}(y)}}{\sqrt{2}} \left(\begin{array}{c} 1 \\ -\ii \ee^{\ii \theta_{1/2}(y)} \end{array}\right)\,,
\end{equation}
%
%
where $k_\text{F}$ is the Fermi momentum of the wave associated to the Fermi energy $\mathcal{E}_\text{F}=\hbar v_\text{F} k_\text{F}$, and $r_i$ are the path the waves are propagating before interfering, see Eq.~\eqref{paths}.  
As for the SE case, the expression for the interference as a function of the position $y$ along the detection plane is given by 
%
%
\begin{eqnarray}\label{interference}
\mathcal{I}_\text{SML}(y) & = &|\tilde{\Psi}_1+\tilde{\Psi}_2|^2 \nonumber \\
 & = & 2+\cos\left[k_\text{F}\left(r_1-r_2\right)\right] \nonumber \\
& & + \cos\left[k_\text{F}\left(r_1-r_2\right)+\theta_1-\theta_2\right].
\end{eqnarray}
%
%
By introducing $\Delta\theta=\theta_1-\theta_2$, we can simplify the previous expression in a more insightful form:
%
%
\begin{eqnarray}\label{int_SML}
\mathcal{I}_\text{SML}(y) &=& 2+\cos(k_\text{F}\Delta r  )[1+\cos(\Delta\theta)] \nonumber \\
&& -\sin(k_\text{F}\Delta r)\sin\Delta\theta.
\end{eqnarray}
%
%
Here we observe that for $\Delta\theta=0$ the correction due to SML goes to zero, and the interference pattern reduces to the SE case in Eq.~\eqref{int_SE}, whereas it is maximum for $\Delta\theta=\pm\pi/2$ and it is equal to $\mathcal{I}_\text{SML}(P)=2+\cos(k_\text{F}\Delta r)-\sin(k_\text{F}\Delta r)$. 
%
%
\begin{figure}[!tb]
\begin{center}
\includegraphics[width=\linewidth]{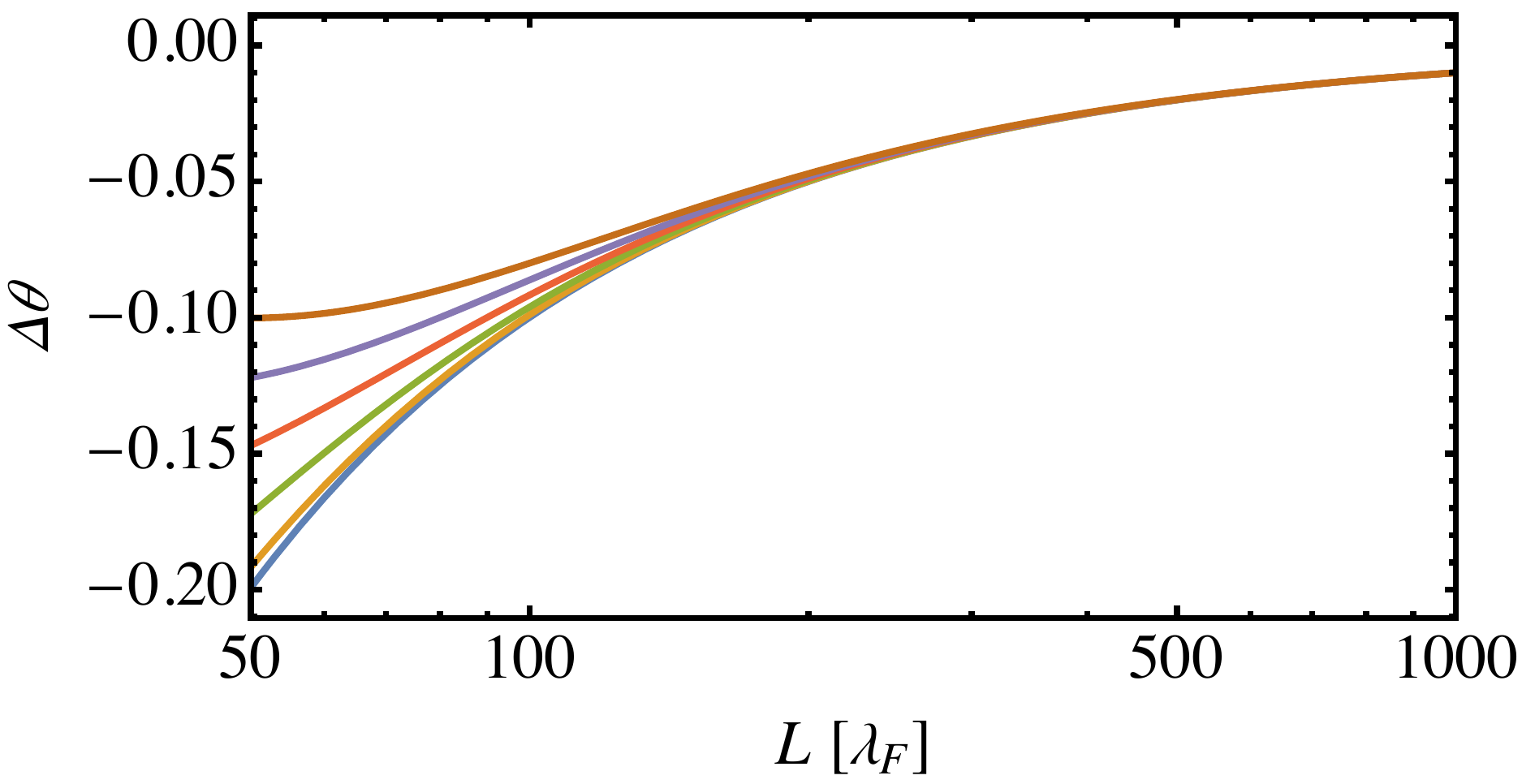}
\caption{\label{fig:Deltatheta} The shift $\Delta\theta$ introduced by SML as a function of the position of the measurement plane $L$ for various detection points $y$ with respect to the center of the system. The various values of $y$ from  bottom to the top are: $0,10,20,30,40,50~\lambda_\text{F}$. We have fixed $d=10~\lambda_\text{F}$. The plot shows the crossover from near-field to far-field where the corrections $\Delta\theta$ becomes negligible.}
\end{center}
\end{figure}
%
%

The expression of the spin angle $\theta_i(x)$ can be obtained by a trigonometric arguments and reads:
%
%
\begin{equation}\label{theta}
\theta_{1/2}(y)= \arctan\left( \frac{y\mp\frac{d}{2}}{L}\right)\,.
\end{equation}
%
%
The difference between the two angles can be expressed as
%
%
\begin{subequations}\label{thetadiff}
\begin{eqnarray}
\Delta\theta &=&\theta_1-\theta_2 \nonumber\\
&=& \arctan\left(\frac{y-\frac{d}{2}}{L}\right)-\arctan\left(\frac{y+\frac{d}{2}}{L}\right)\label{thetadiff_a}\\
&=& \arctan\left[\frac{4dL}{d^2-4(L^2-y^2)}\right]\,, \label{thetadiff_b}
\end{eqnarray}
\end{subequations}
%
%
that is an even function of the detector position $y$. The behavior of this function for fixed $d$, and as a function of $L$ for various position of the detector $y$, is shown in Fig.~\ref{fig:Deltatheta}. This angle difference is zero for the two asymptotic values $y\to \pm \infty$ and is largest and negative for $y\to0$ with $\Delta\theta(0)=-2\arctan\left[d/2L\right]$. As mentioned above, the first limit can be easily understood because for measurement points far away from the center of the system, the two electrons arrive with almost the same propagation direction, thus with parallel spins; on the contrary, the second limit corresponds to maximizing the orientation difference of the two spins. We learn from Eq.~\eqref{thetadiff} that the correction to the double slit pattern due to SML depends only on  the geometrical parameters of the system $y$, $d$ and $L$, whereas it does not depend on Fermi wavelength $\lambda_\text{F}$.

%
\begin{figure}[!tb]
\begin{center}
\includegraphics[width=\columnwidth]{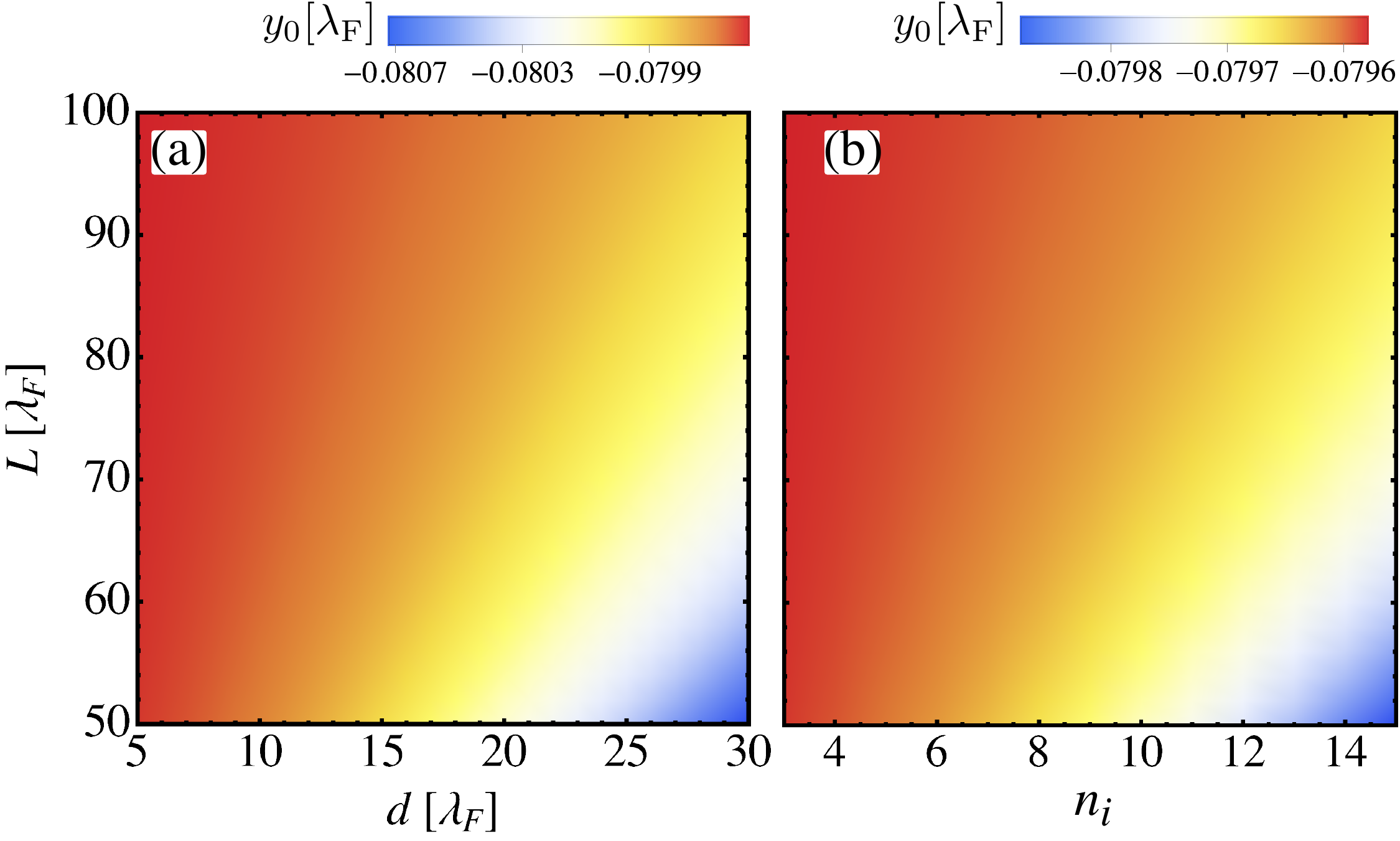}
\caption{\label{fig:shift:both} (a) Shift of the first maximum on the interference for the case with SML as a function of the distance between the slits $d$ and the distance of the measurement plane $L$; (b) Shift of the first maximum on the diffraction for the case with SML as a function of the number of secondary sources in the slit $n_i$ and the distance of the measurement plane $L$. In both panels, the shift is expressed in units of the Fermi wavelength $\lambda_\text{F}$.}
\end{center}
\end{figure}
%
%
Interestingly, the main effect on the interference pattern compared to the SE case is to produce a shift of the position $y_0$ of the first maximum, see Fig.~\ref{fig:shift:both}. The shift depends on the various geometrical parameters, namely  the slit distance $d$ and the distance of the observation plane $L$. In Fig.~\ref{fig:shift:both}(a) we show how $y_0$ depends on these parameters: we clearly observe that the shift is very small in terms of the Fermi wavelength $\lambda_\text{F}$. 

\subsubsection{Diffraction from a single slit}

As we already did for the case of spinless electrons, we now relax the condition $h\sim\lambda_\text{F}$, to study the modifications of the diffraction pattern due to the SML.
Under this assumption, the global wavefunction at the slit $\tilde{\Psi}_\text{tot}^\text{dif}$ can be written as:
%
%
\begin{equation}
    \tilde{\Psi}_\text{tot}^\text{dif}=\frac{1}{\sqrt{2}}\sum_{m=1}^N \ee^{\ii k_\text{F}r_m}\begin{pmatrix}1\\ -\ii \ee^{\ii \theta_m} \end{pmatrix}\,,
\end{equation}
%
%
where $r_m$ is the path between the emitter and the observation point P, and $\theta_m$ is the associated angle of propagation that determines the spin direction imposed by SML. The total intensity $\mathcal{D}_\text{SE}$ at the observation point P is given by
%
%
\begin{eqnarray}\label{diff_SML}
    \mathcal{D}_\text{SML}(y)&=&|\tilde{\Psi}_\text{tot}^\text{dif}|^2 \nonumber \\
    &=&N+\sum_{\substack{m,n=1\\n\neq m}}^N \left\{\cos\left[k_\text{F}(r_n-r_m)\right]\right. \nonumber \\
    &&\hspace{0.4cm}\left.+\cos\left[k_\text{F}(r_n-r_m)+(\theta_n-\theta_m)\right]\right\},
\end{eqnarray}
%
%
In this expression for the diffraction in the presence of SML, the angles due to the spinorial structures of the wave function are obtained by a generalization of the Eq.~\eqref{theta}, with the substitution of the width of the opening $d$ with the distance $a$ between the electron emitters --- see Fig.~\ref{fig:young}(b). It is clear that in absence of SML, the expression for the diffraction pattern in Eq.~\eqref{diff_SML} reduces to the one in the case of SE in Eq.~\eqref{diff_SE}. 

Here again, we see the main effect of the SML is a shift of the position $y_0$ of the first maximum of the diffraction pattern, which is depicted in Fig.~\ref{fig:shift:both}(b) as a function of the number of secondary emitters $n_i$ within the slit, and the distance of the screen $L$.

As we did for the SEs in Eq.~\eqref{far_field_diff_SE}, we can for completeness recast the expression into a more compact form in the far-field limit, where $k_\text{F}(r_n-r_m)\to (n-m)\varphi$ and $\theta_m-\theta_n\to (n-m)\chi$; in this case Eq.~\eqref{diff_SML} reduces to:
%
%
\begin{equation}\label{far_field_diff_SML}
    \mathcal{D}_\text{SML}= \frac{1}{2}\left[ \frac{\sin\left(N \frac{\varphi}{2}\right)}{\sin\left(\frac{\varphi}{2}\right)}\right]^2+\frac{1}{2}\left[ \frac{\sin\left(N \frac{\varphi+\chi}{2}\right)}{\sin\left(\frac{\varphi+\chi}{2}\right)}\right]^2\,
\end{equation}
%
%
where we defined $\varphi$ in Eq.~\eqref{defvarphi}, and the shift due to the SML is of geometric origin only, and given by
%
%
\begin{equation}\label{defchi}
    \chi=\frac{a}{L}.
\end{equation}
%
%

\subsection{High-energy case for the 3DTI}\label{section_high_energy}

\begin{figure}
    \centering
    \includegraphics[width=\columnwidth]{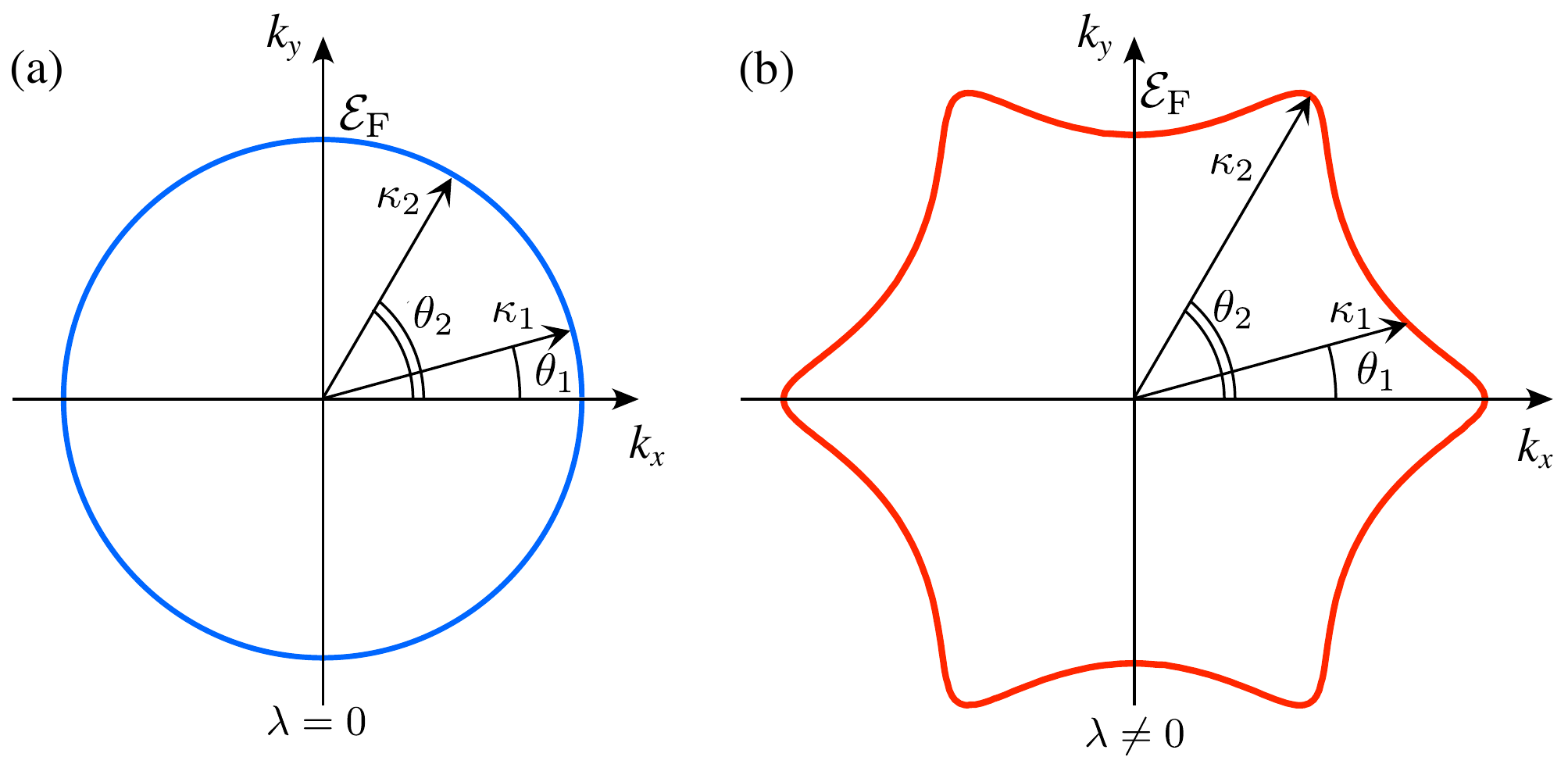}
    \caption{\label{fig_fermi_energy} Fermi surfaces for the low-energy case (a) and the high-energy one (b). The energy is fixed to $\mathcal{E}_\text{F}$.}
    
\end{figure}

In this section, we consider a Hamiltonian description of the surface states beyond the low-energy approximation introduced in Eq.~\eqref{eq_low_energy}. It was observed experimentally that the shape of the Fermi surface describing the surface states of 3DTIs in layered materials, such as Bi$_2$Te$_3$, is circular only close to the crossing point of the Dirac dispersion. At higher energies it departs from this shape, becoming at first hexagonal and then assuming a snowflake shape~\cite{Chen_2009}. These changes in the structure of the Fermi surface can be accounted for within $k \cdot p$ theory by adding an additional spin-orbit term to the Hamiltonian~\eqref{eq_low_energy} describing the hexagonal warping~\cite{Fu_2009}. This term can be constructed for surface states of rhombohedral crystalline structures solely from symmetry considerations. It can reproduce the results obtained by ab-initio calculations, however its simple structure allows for a simplified analysis revealing basic properties of topological edge states. Within this framework, the effective $k \cdot p$ Hamiltonian describing the surface states of a 3DTI reads: 
%
%
\begin{equation}\label{eq_warping}
    \mathcal{H}_\text{3DTI+w}=\mathcal{H}_\text{3DTI}+\frac{\beta}{2}\left(\bm{p}_+^3+\bm{p}_-^3\right)\sigma_z,
\end{equation}
%
%
where $\beta$ is the strength of the hexagonal warping and $\bm{p}_\pm=\bm{p}_x\pm \ii \bm{p}_y$. A more in-depth analysis of the $k\cdot p$ Hamiltonian can be found in  Ref.~\cite{Fu_2009}. \\
This Hamiltonian is characterized by the following spectrum:
%
%
\begin{equation}\label{spectrum_he}
    \mathcal{E}(\kappa,\theta)=\pm\hbar\sqrt{v_\text{F}^2 \kappa^2+\beta^2\kappa^6 \cos^2(3\theta)}
\end{equation}
%
%
where $\kappa$ is the modulus of the momentum, and the motion direction is characterized by the azimuthal angle already defined earlier as $\theta=\arctan(k_y/k_x)$. We learn from this expression of the spectrum at high energy that the states at the Fermi energy $\mathcal{E}_\text{F}$ can have different values of the modulus of the momentum $\kappa_\text{F}$ for different propagation directions $\theta$. This is different from the circular symmetry configuration we have in the low-energy limit, for which the modulus of the momentum is the same for all the propagation directions --- see the two panels of Fig.~\ref{fig_fermi_energy}.  
The eigenstates maintain a spinorial structure that reads:
%
%
\begin{equation}\label{vectors_hw}
    v_\pm(\kappa,\theta) =\frac{1} {\sqrt{2\mathcal{N}}} \begin{pmatrix}
    b_\beta^\mp(\kappa,\theta) \\ \ii \ee^{\ii \theta}
    \end{pmatrix},
\end{equation}
%
%
where the normalization factor for the spinor is defined as
%
%
\begin{equation}\label{normalization}
    2\mathcal{N}=[b_\beta^-(\kappa,\theta)]^2+1,
\end{equation}
%
%
and the functions $b_\beta^\pm(\kappa,\theta)$ are  defined as
%
%
\begin{equation}\label{b_function}
    b_\beta^\pm(\kappa,\theta) = \kappa^2 \beta \cos(3\theta)\pm\sqrt{1+v_\text{F}^{-2}\kappa^4 \beta^2 \cos^2(3\theta)}.
\end{equation}
%
%
When the warping is absent ($\beta\to0$) the states in Eq.~\eqref{vectors_hw} coincide with the spinors in the low-energy approximation in Eq.~\eqref{eq_ei_SML_LE}.
%
%

%
%

\subsubsection{Interference from double-slit set-up}

As we already did in the low-energy approximation, we can define a wave-function at positive energy $\mathcal{E}_\text F $ and fixed propagation direction $\theta$ as:
%
%
\begin{equation} \label{wavesintHE}
\bar{\Psi}_{1/2}(y)=\frac{\ee^{\ii \kappa_{1/2} r_{1/2}(y)}}{\sqrt{2 \mathcal{N}_{1/2}}} \begin{pmatrix} 
b_\beta^-(\kappa_{1/2},\theta_{1/2}) \\ \ii \ee^{\ii \theta_{1/2}(y)} 
\end{pmatrix},
\end{equation}
%
%
where $\kappa_{1/2}$ are the modulus of the momenta associated with the propagation directions $\theta_{1/2}$, respectively (see Fig.~\ref{fig_fermi_energy}); proceeding as we did in the previous section, we find the following expression for the interference pattern:
%
%
\begin{align}\label{intHE}
\bar{\mathcal{I}}_\text{SML}= 2 &+\frac{1}{\sqrt{\mathcal{N}_1\mathcal{N}_2}}\big[\cos(\zeta_{1/2}) b_\beta^-(\kappa_1,\theta_1) b_\beta^-(\kappa_2,\theta_2) \nonumber\\
    & + \cos(\zeta_{1/2} +\Delta \theta)\big]\,,
\end{align}
%
%
where $\zeta_{1/2} = \kappa_1 r_1 - \kappa_2 r_2$. Importantly, we note from this last term that the phase difference for the electron arriving from the two slits in addition to depending on the different geometric paths $r_1$ and $r_2$ also depends on the different momenta $\kappa_1$ and $\kappa_2$. This is due to the distortion of the Fermi surface caused by the warping in Hamiltonian~\eqref{eq_warping} and can be considered as an additional signature of the SML. In Fig.~\ref{fig_shift_he}, we present the shift of the position of the first maximum of interference obtained by Eq.~\eqref{intHE} as a function of the distance between the slits $d$ and the position of the measurement plane $L$. We analyse the shift for two different values of the warping of the Fermi surface. The value of the shift and the behaviour with the parameters is different for the two values of $\beta$.

\subsubsection{Diffraction from a single slit}
\begin{figure}[!t]
    \centering
    \includegraphics[width=0.8\columnwidth]{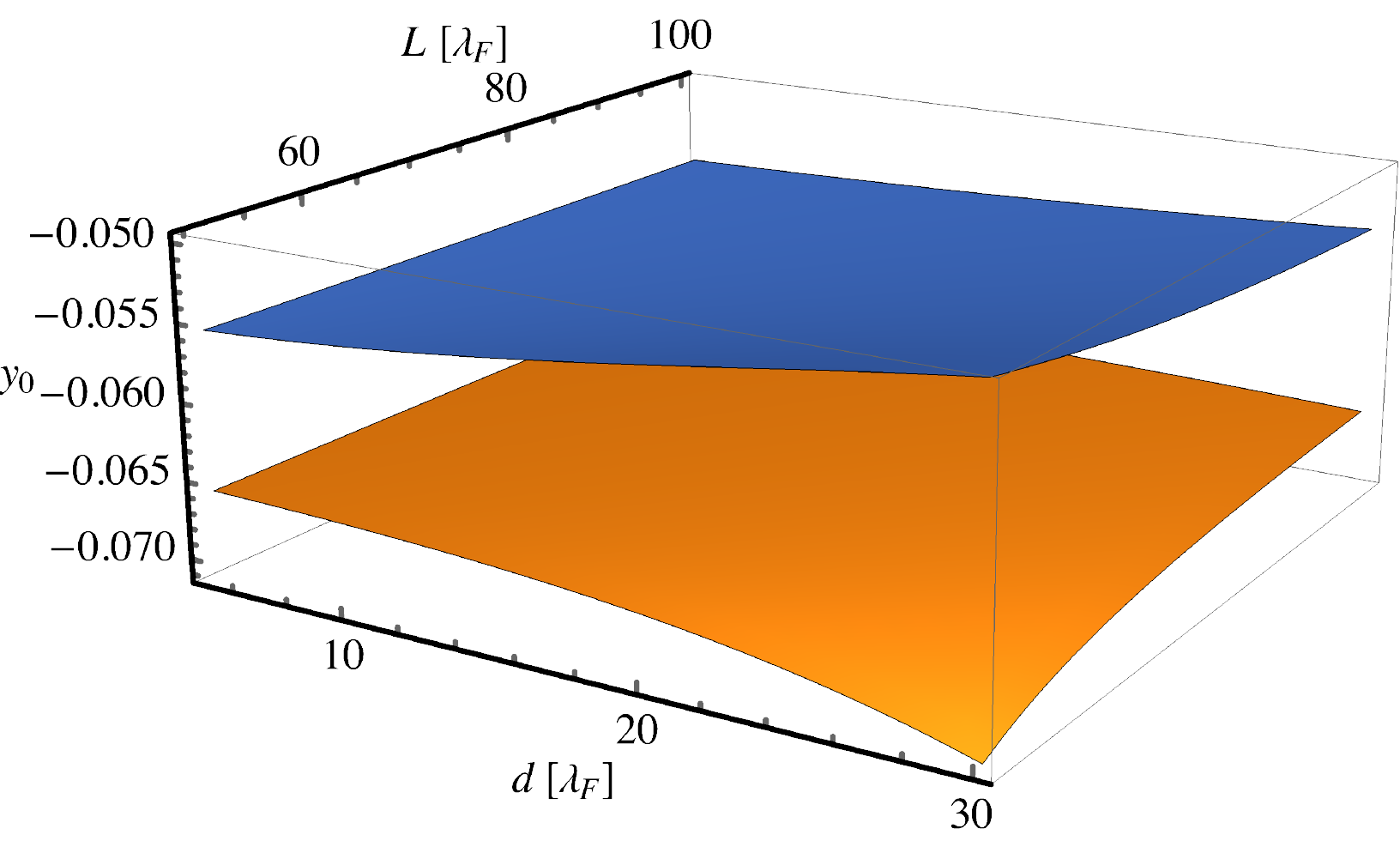}
    \caption{Shift of the position  first maximum of interference $y_0$ for finite hexagonal warping: (a) $\beta=0.1~v_\text{F}$ and (b) $\beta=0.2~v_\text{F}$.}
    \label{fig_shift_he}
\end{figure}

In order to define the diffraction pattern in the high-energy limit, we introduce the following wave function accounting for $N$ secondary emitters in the slit of opening $h$:
%
%
\begin{align}
    \bar{\Psi}^\text{dif}_\text{tot}=  \sum_{m=1}^N \frac{\ee^{\ii \kappa_m r_m}}{\sqrt{2\mathcal{N}_m}}
    \begin{pmatrix}
        b_\beta^-(\kappa_m,\theta_m) \\
        \ii \ee^{\ii \theta_m}
    \end{pmatrix},
\end{align}
%
%
The total intensity $\bar{\mathcal{D}}_\text{SML}$ at the observation point P is given by
%
%
\begin{align}\label{diff_SML_HE}
    &\bar{\mathcal{D}}_\text{SML}(y)=|\tilde{\Psi}_\text{tot}^\text{dif}|^2 \nonumber \\
    &= N+\sum_{\substack{n,m=1\\ n\neq m}}^N \frac{1}{\sqrt{\mathcal{N}_n\mathcal{N}_m}}\big\{ b_\beta^-(\kappa_m,\theta_m)b_\beta^-(\kappa_n,\theta_n)\cos\zeta_{n/m} \nonumber \\ &\hspace{0.8cm}+\cos\left[\zeta_{n/m}+(\theta_n-\theta_m)\right]\big\},
\end{align}
%
%
where $\zeta_{n/m}=\kappa_n r_n-\kappa_m r_m$ and the normalization constants are  $\mathcal{N}_m=[b_\beta^-(\kappa_m,\theta_m)]^2+1$. This expression, in the limit of zero warping ($\beta\to0$), is identical to the diffraction expression in Eq.~\eqref{diff_SML}. Also in this case we observe a shift of the first maximum of diffraction that behaves similarly to the interference case we have shown in Fig.~\ref{fig_shift_he}, as can be seen from Eq.~\eqref{diff_SML_HE} (not plotted).
%
%
%
%
\section{Comparisons}\label{comparison}
%
%
\begin{figure}[!tb]
\begin{center}
\includegraphics[width=\linewidth]{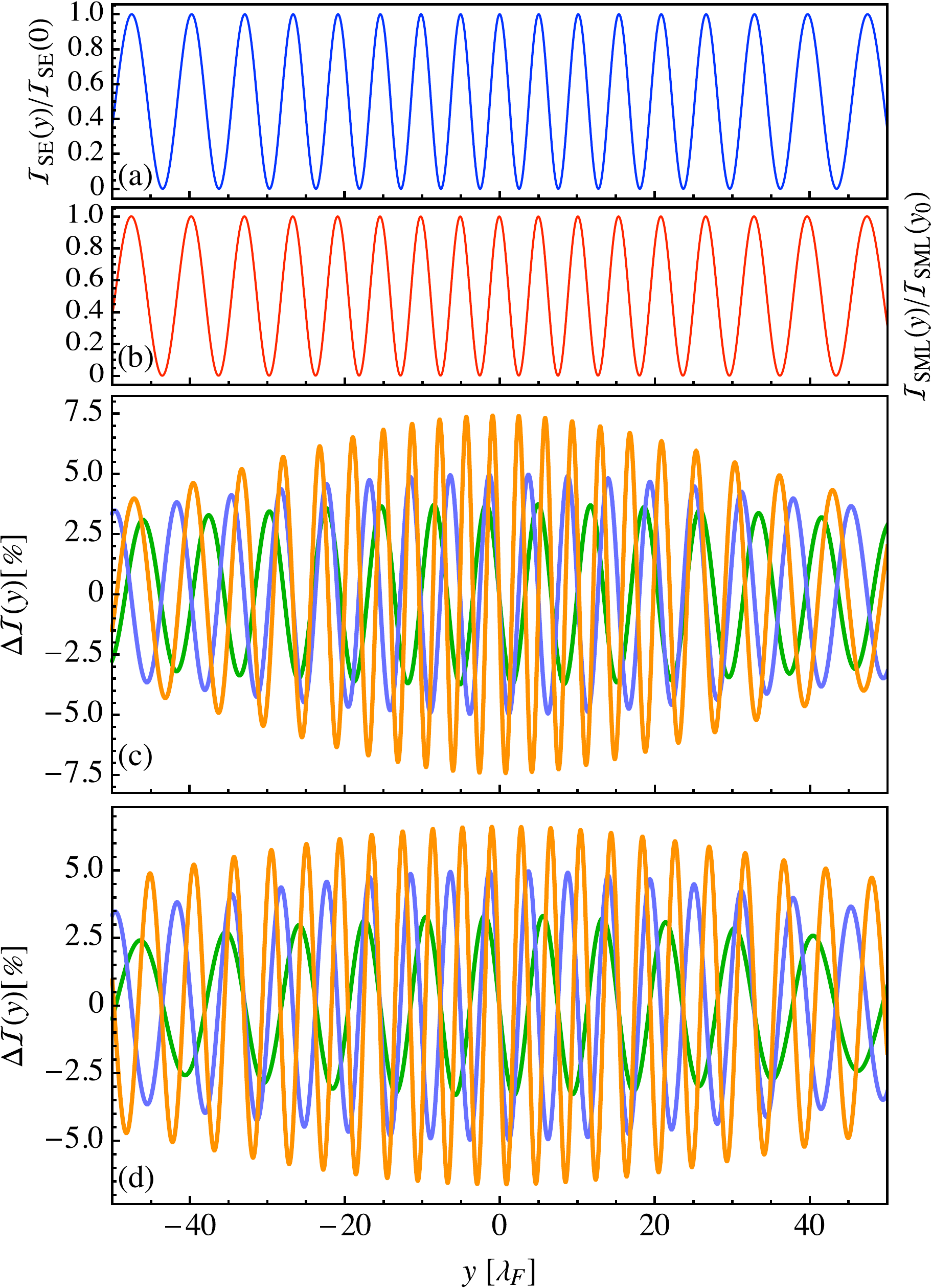}
\caption{\label{fig:comparison:int} Interference pattern from a double-slit system as a function of the detection point $y$: (a) case of spinless electrons; (b) case of electrons with spin-momentum locking in the low-energy approximation. For panel (a) and (b), we have $d=15~\lambda_\text{F}$ and $L=75~\lambda_\text{F}$. Panels (c) and (d) give the percentage difference of the two previous cases for $100 \Delta\mathcal{I}(y)$ in Eq.~\eqref{eq_comparison}. In panel (c), the distance between the slits is kept constant at $d=15~\lambda_\text{F}$ and the distance of the $L$ is varied between $50~\lambda_\text{F}$ (orange), $75~\lambda_\text{F}$ (blue), $100~\lambda_\text{F}$ (green). In panel (d), the distance between of the measurement plane is kept constant at $L=75~\lambda_\text{F}$ and the distance of the $d$ is varied between $10~\lambda_\text{F}$ (green), $15~\lambda_\text{F}$ (blue), $20~\lambda_\text{F}$ (orange).}
\end{center}
\end{figure}
%
%
First, we will compare the case of SE with that of SML at low-energy. In this case, we can define in a unique way a Fermi wavelength, the interference pattern will be characterized by a single oscillation pattern, due to the absence of distortion of the Fermi surface. For a fixed $\lambda_\text{F}$, the interference patterns in the two cases are almost identical: the oscillation frequency is lead by the Fermi wavelength, but we will observe a shift $y_0$ of the first maximum for the SML case. For this reason, it is more informative to look at the normalized difference between the interference patterns in the SE and SML cases. For this purpose, we define the following function
%
%
\begin{equation}\label{eq_comparison}
    \Delta \mathcal{I}(y)=\frac{\mathcal{I}_\text{SML}(y)-\mathcal{I}_\text{SE}(y)}{\mathcal{I}_\text{SE}(0)}.
\end{equation}
%
%
In Fig.~\ref{fig:comparison:int}(a) and~\ref{fig:comparison:int}(b), we present the interference pattern for the SE and SML cases for a fixed set of parameters $d$ and $L$. In Fig.~\ref{fig:comparison:int}(c) and~\ref{fig:comparison:int}(d) of Fig.~\ref{fig:comparison:int}, we present $\Delta \mathcal{I}(y)$ for fixed $d$ and different $L$, and for fixed $L$ and various $d$, respectively. We see from Fig.~~\ref{fig:comparison:int}(c) that this difference can be of the order of almost $7.5\%$. It is important to note that in the two interference figures in Fig.~\ref{fig:comparison:int}(a) and~\ref{fig:comparison:int}(b), a distortion of the oscillation pattern can be observed for large values of the position $y$. This distortion is present both in the SE and SML cases, this is a consequence of not considering the far-field limit in the position function given in Eq.~\eqref{deltar}.

%
%
\begin{figure}[!t]
\begin{center}
\includegraphics[width=\linewidth]{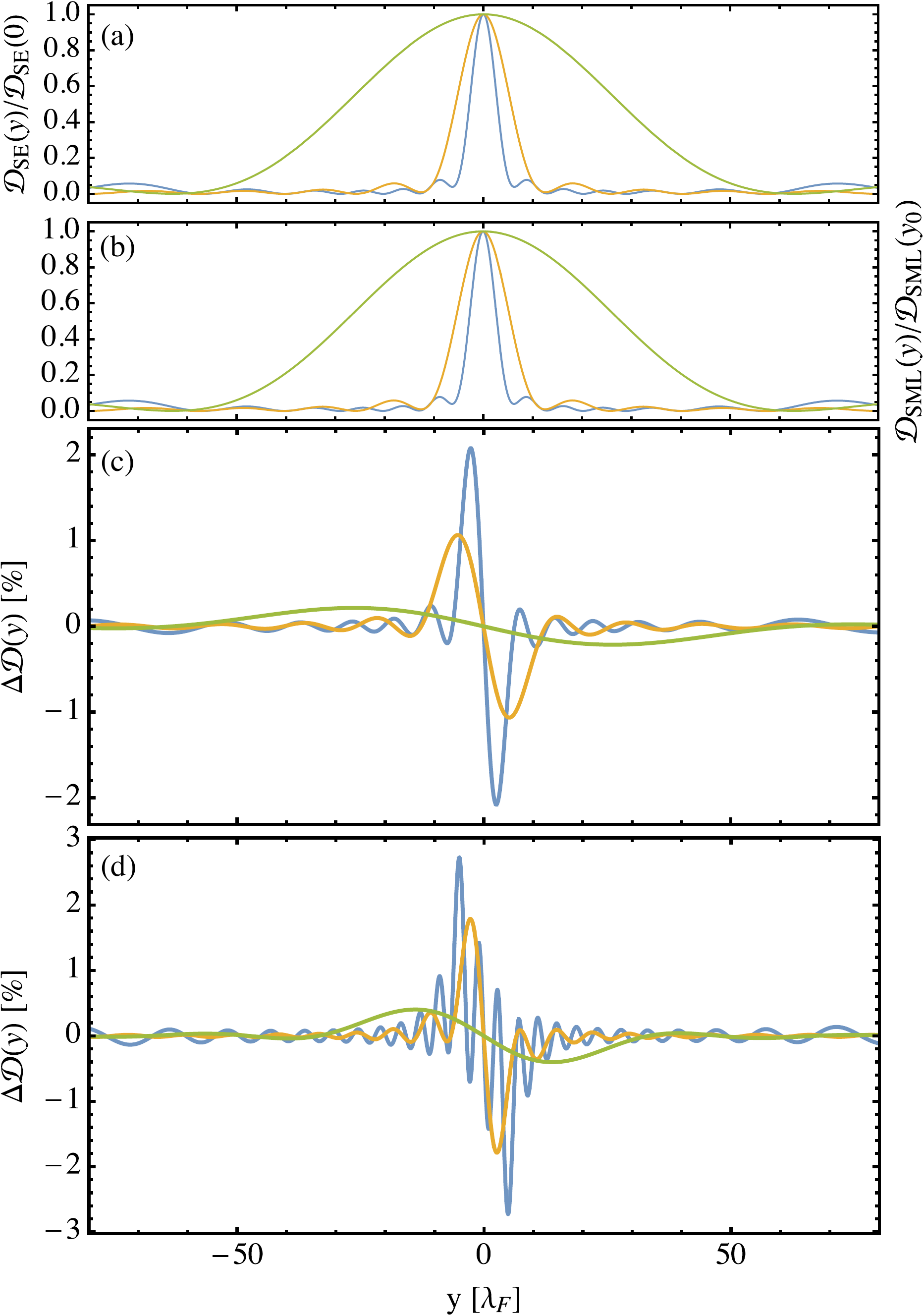}
\caption{\label{fig:comparison:diff} Diffraction pattern from a single-slit system as a function of the detection point $y$: (a) case of spinless electrons; (b) case of electrons with spin-momentum locking in the low-energy approximation. For panel (a) and (b), we have $L=50~\lambda_\text{F}$ (blue), $L=100~\lambda_\text{F}$ (yellow) $L=500~\lambda_\text{F}$ (green), the number of secondary emitter is set to $8$. Panel (c) and (d) percentage difference of the two previous case $100 \Delta\mathcal{D}(y)$ in Eq.~\eqref{eq_comparison_dif}. In panel (c) we consider the case of $8$ secondary emitters and the three distance $L$ as in panel (a) and (b), in panel (d) the number of secondary emitters is $15$ and the three distance $L$ as in panel (a) and (b).}
\end{center}
\end{figure}
%
%
We now perform the same kind of comparison for the case of the single-slit diffraction pattern; as for the interference case, we introduce an intensity ratio function:
%
%
\begin{equation}\label{eq_comparison_dif}
    \Delta \mathcal{D}(y)=\frac{\mathcal{D}_\text{SML}(y)-\mathcal{D}_\text{SE}(y)}{\mathcal{D}_\text{SE}(y_0)}
\end{equation}
%
%
%
for a fixed set of geometrical parameters $h$ and $L$. The comparison of the diffraction pattern for various sizes of the single slit $h$ is shown in Fig.~\ref{fig:comparison:diff} moving from the near-field to the far-field case. For an observation plane placed at larger distance, we observe a larger central maximum whereas the side maxima are not always visible. Similarly to the double-slit case, we observe that the corrections [Fig.~\ref{fig:comparison:diff}(c) and (d)] are larger when the observation plane is placed at a distance $L$ not too large compared to the overall opening of the slit $h$. The corrections can be up to the order of $2.5\%$ in the case of a larger slit in Fig.~\ref{fig:comparison:diff}(d).

%
%
\begin{figure}[!tb]
\begin{center}
\includegraphics[width=\linewidth]{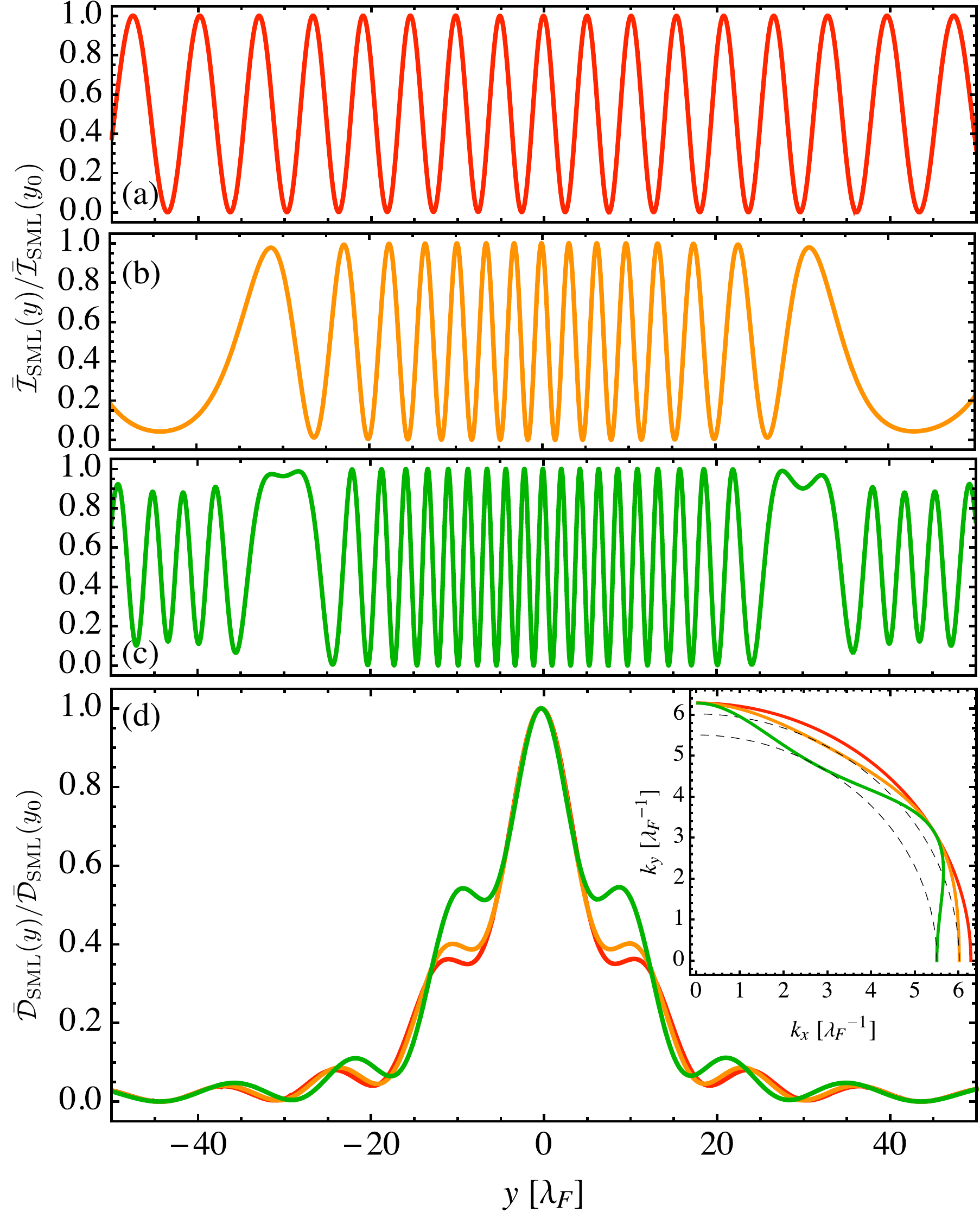}
\caption{\label{fig_high_energy} The warping effects from Eq.~\eqref{eq_warping}, the interference pattern from a double-slit is shown in panel (a) to (c) for $L=75~\lambda_\text{F}$ and $d=15~\lambda_\text{F}$ and $\mathcal{E}_\text{F}=2\pi$, in panel (a) there is no warping, in panel (b) $\beta=0.05$ and in panel (c) $\beta=0.1$. The diffraction pattern from a single slit is shown in panel (d) for the same energy as in (a) with $8$ secondary emitters and $L=75~\lambda_\text{F}$, the three curves correspond to the same value of $\beta$ as in panel (a) to (c). In the inset of panel (d), we show the Fermi surface corresponding to the two values of $\beta$ and $\mathcal{E}_\text{F}=2\pi$.}
\end{center}
\end{figure}
%
%
We will now present results of the interference pattern from a double-slit and the diffraction pattern from a single-slit, for the case with warping corrections as in Hamiltonian~\eqref{eq_warping}. We will compare these results to the SML in the low-energy case with $\beta=0$, other comparisons are problematic due to the impossibility of defining a unique Fermi wavelength in the high-energy case, as discussed in Sec.~\ref{section_high_energy}. We present two different cases with a finite warping $\beta$ for the interference pattern in Fig.~\ref{fig_high_energy}(b) and~\ref{fig_high_energy}(c). We immediately observe a second period of oscillation due to the different momenta characterizing the Fermi surface in the presence of warping --- see inset of Fig.~\ref{fig_high_energy}(d). The main effect in the diffraction pattern from a single-slit, in addition to introducing a change in the oscillations, is to enlarge the size of the central maximum.

\section{Conclusions and outlook}\label{conclusions}

In this article we have evaluated the effect of the locking of the spin-degree-of-freedom to the motion direction, typical for systems with spin-orbit interaction. We have shown that the spin-momentum locking leads to a small but finite correction to the interference and diffraction patterns for electrons going through a single or a double-slit system. In particular, we have investigated a two-dimensional electron system given by the electrons confined on the surface of three-dimensional topological insulators. This electron system has the peculiarity of having a kinetic Hamiltonian that is originating from spin-orbit interaction only. We considered both the low-energy limit and the high-energy generalization, in which case a cubic in momentum spin-orbit term takes into account the warping of the Fermi energy. In the high-energy limit we found, in addition to a small correction, the appearance of different oscillatory contributions arising from the warping of the Fermi surface.

Here, we present a proposal for realizing a slit experiment in a solid-state platform based on using the two-dimensional electron gas of noble metals for the SE case, and a 3DTI for the case of SML electrons. Both require the use of an STM for the manipulation of atoms of a surface. In the SE case, we can consider CO molecules placed on Cu(111)~\cite{Gomes_2012,Slot_2017,Kempkes_2018,Kempkes_2019}, this technique allows to place with atomic precision molecules on specific positions of the metal surface; in general, the CO molecules act as repulsive barriers for the free electrons on the surface of the noble metal. The CO molecules can be arranged to form two parallel walls, one containing one or more slits and the opposite without opening. Similar to the case of a terrace or step edges on the surface Cu(111) the two walls will generate standing plane waves~\cite{Davis_1991,Crommie_1993,Hasegawa_1993}. The tip of the STM can then be used as a local probe for measuring the local density of states at the detection point P on the screen --- see Fig.~\ref{fig:young}(a). In a similar fashion, we can measure the shift introduced by the SML, by substituting the Cu(111) surface with a 3DTI material such as Bi$_2$Te$_3$ or Bi$_2$Se$_3$. We then also need to substitute the CO molecules with magnetic molecules~\cite{Sessi_2016}, so as to locally gap the linear dispersion --- see Eq.~\eqref{ham:3DTI}. For the case of a 3DTI, a weak magnetic field can also be used to enhance some of the physics discussed in this manuscript. It should lead to an additional shift of the first maximum of interference for the case of SML, while it will have no significant effect in the case of SEs.

\begin{acknowledgements}
We acknowledge useful discussions with Alessandro De Martino, Erwann Bocquillon, M. Reyes Calvo, Geza Giedke, Andreas Inhofer, and Ingmar Swart. The work of TB and DB is supported by the Spanish Ministerio de Ciencia, Innovation y Universidades (MICINN) through the Project FIS2017-82804-P, and by the Transnational Common Laboratory \emph{Quantum-ChemPhys}.
DB thanks the University of Aix-Marseille for hosting him during the genesis of this work. 
\end{acknowledgements}
\appendix
\section{On the diffraction formula}

In this appending we present the derivation for the diffraction formula in the leading order in $L^{-1}$ presented in Eq.~\eqref{far_field_diff_SE} or \eqref{far_field_diff_SML}. The main step is to expand all the $\cos[(n-m)\varphi]=\cos\ell\varphi$ into exponential functions so that
%
%
\begin{subequations}
\begin{align}
    \mathcal{D}_\text{SE} &\longrightarrow \ee^{-\ii (N-1)\varphi} \left(\sum_{m=0}^N \ee^{\ii m \varphi}\right)^2\!,\label{step_one} \\
    &=\ee^{-\ii (N-1)\varphi} \left( \frac{1-\ee^{\ii N \varphi}}{1-\ee^{\ii\varphi}}\right)^2\!, \label{step_two} \\
    &= \ee^{\ii (N-1)\varphi}\left(\ee^{\ii\frac{(N-1)\varphi}{2}}\frac{\sin(N\varphi/2)}{\sin(\varphi/2)}\right)^2\!,\label{step_three} \\
    &= \left[\frac{\sin\left(\frac{N\varphi}{2}\right)}{\sin\left(\frac{\varphi}{2}\right)}\right]^2\!,\label{step_four}
    \end{align}
\end{subequations}
%
%
where in \eqref{step_one} we have used the properties of geometric series: $\sum_{n=0}^N r^n=(1-r^{N+1})/(1-r)$. A similar expression is obtained starting by Eq.~\eqref{diff_SML} with the introduction of the spin parameter $\chi$. 

\bibliographystyle{spphys}
\bibliography{Bibliography}

\end{document}